\documentclass[%
reprint,
twocolumn,
notitlepage,
superscriptaddress,
jmp,%
amsmath,amssymb,
 pra,
letter,%
]{revtex4-2}
\usepackage{graphicx}
\usepackage{subfigure}
\usepackage{dcolumn}
\usepackage{bm}
\usepackage[mathlines]{lineno}
\graphicspath{{figure/}}
\usepackage{array}
\usepackage{braket}
\usepackage{diagbox}
\usepackage{booktabs}
\usepackage{textcomp}
\usepackage{amsmath}
\usepackage{appendix}
\usepackage{multirow}
\usepackage{bm,color,bbm}

\usepackage{float}
\newcolumntype{.}{D{.}{.}{-1}}
\usepackage{amsopn}
\usepackage[colorlinks,linkcolor=blue, anchorcolor=blue, urlcolor=blue, citecolor=blue]{hyperref}
\usepackage{epstopdf}

\begin{document}
\title{Silicon-based decoder for polarization-encoding quantum key distribution}

\author{Yongqiang Du}
\thanks{Equal contribution}
\affiliation  {Guangxi Key Laboratory for Relativistic Astrophysics, School of Physical Science and Technology, Guangxi University, Nanning 530004, China}

\author{Xun Zhu}
\thanks{Equal contribution}
\affiliation {National Information Optoelectronics Innovation Center (NOEIC), 430074, Wuhan, China}

\author{Xin Hua}
\affiliation {National Information Optoelectronics Innovation Center (NOEIC), 430074, Wuhan, China}
\affiliation {State Key Laboratory of Optical Communication Technologies and Networks, China Information and Communication Technologies Group Corporation (CICT), 430074, Wuhan, China}
\author{Zhengeng Zhao}
\affiliation  {Guangxi Key Laboratory for Relativistic Astrophysics, School of Physical Science and Technology, Guangxi University, Nanning 530004, China}
\author{Xiao Hu}
\affiliation {National Information Optoelectronics Innovation Center (NOEIC), 430074, Wuhan, China}
\affiliation {State Key Laboratory of Optical Communication Technologies and Networks, China Information and Communication Technologies Group Corporation (CICT), 430074, Wuhan, China}
\author{Yi Qian}
\affiliation {National Information Optoelectronics Innovation Center (NOEIC), 430074, Wuhan, China}
\affiliation {State Key Laboratory of Optical Communication Technologies and Networks, China Information and Communication Technologies Group Corporation (CICT), 430074, Wuhan, China}
\author{Xi Xiao}
\thanks{Corresponding author: xxiao@wri.com.cn}
\affiliation {National Information Optoelectronics Innovation Center (NOEIC), 430074, Wuhan, China}
\affiliation {State Key Laboratory of Optical Communication Technologies and Networks, China Information and Communication Technologies Group Corporation (CICT), 430074, Wuhan, China}
\author{Kejin Wei}
\thanks{Corresponding author: kjwei@gxu.edu.cn}
\affiliation  {Guangxi Key Laboratory for Relativistic Astrophysics, School of Physical Science and Technology, Guangxi University, Nanning 530004, China}

\begin{abstract}
	Silicon-based polarization-encoding quantum key distribution (QKD) has been widely studied, owing to its low cost and robustness. However, prior studies have utilized off-chip devices to demodulate the quantum states or perform polarization compensation, given the difficulty of fabricating polarized independent components on the chip. In this paper we propose a fully chip-based decoder for polarization-encoding QKD. The chip realizes a polarization state analyzer and compensates for the BB84 protocol without requiring additional hardware. It is based on a polarization-to-path conversion method that uses a polarization splitter-rotator. The chip was fabricated using a standard silicon photonics foundry; it has a compact design and is suitable for mass production. In the experimental stability test, an average quantum bit error rate of $0.56\%$ was achieved through continuous operation for 10 h without any polarization feedback. Furthermore, using the developed feedback algorithm, the chip enabled the automatic compensation of the fiber polarization drift, which was emulated by a random fiber polarization scrambler. In the case of the QKD demonstration, we obtained a finite-key secret rate of 240 bps over a fiber spool of 100 km. This study represents an important step toward the integrated, practical, and large-scale deployment of QKD systems.
	
\end{abstract}

\maketitle

\section{Introduction}

	Security communication is an indispensable part of government affairs, commerce, national defense, and personal daily life. As the security of traditional public-key cryptography depends on the computational complexity of certain mathematical functions~\cite{PKC1978}, it is threatened gravely by the development of quantum computing science~\cite{zhong2020quantum,arute2019quantum,2022Jin-QC}. The security of quantum key distribution (QKD) is based on the fundamental laws of quantum mechanics and has become a key solution to ensure information security in the information era~\cite{1999Lo}.
	
	Since Bennett and Brassard proposed the first quantum cryptography protocol (BB84) in 1984~\cite{BB84}, QKD experiments have made great progress, and experiments have been performed with different degrees of freedom in photons, such as the time-bin phase~\cite{2018Yuan-time-bin,2017Islam-time-bin,2018-time-bin,2018Wangshuang-time} and polarization~\cite{2018Agnes-pol,2018Gru-polarization} over fiber-based~\cite{Andrew2019,Lo2019TF,Chen2020,wang2022twin,2021-Zhou-xing-RFI}, free space~\cite{2015Vallone-free-space,liao2017satellite,2017Bedinton-free-space}, and underwater~\cite{Feng2021underwater,XianMinJin2019}  channels. QKD networks have also been deployed worldwide~\cite{DARPA-network,SECOQC_network,2011Sasaki-network,2014Wang-NETWORK,19cambridge_network,2021Yang-network}. To achieve intercontinental secure communications, satellites have been used as trusted relays to connect remote user nodes~\cite{Satellite-Relayed}, and large-scale satellite networks~\cite{space} have also been successfully constructed. Furthermore, recent progress in QKD has been observed in the latest research~\cite{xu2020,2020Pirandola}.
	\begin{figure*} [t]
		\centering
		\includegraphics[scale=0.8]{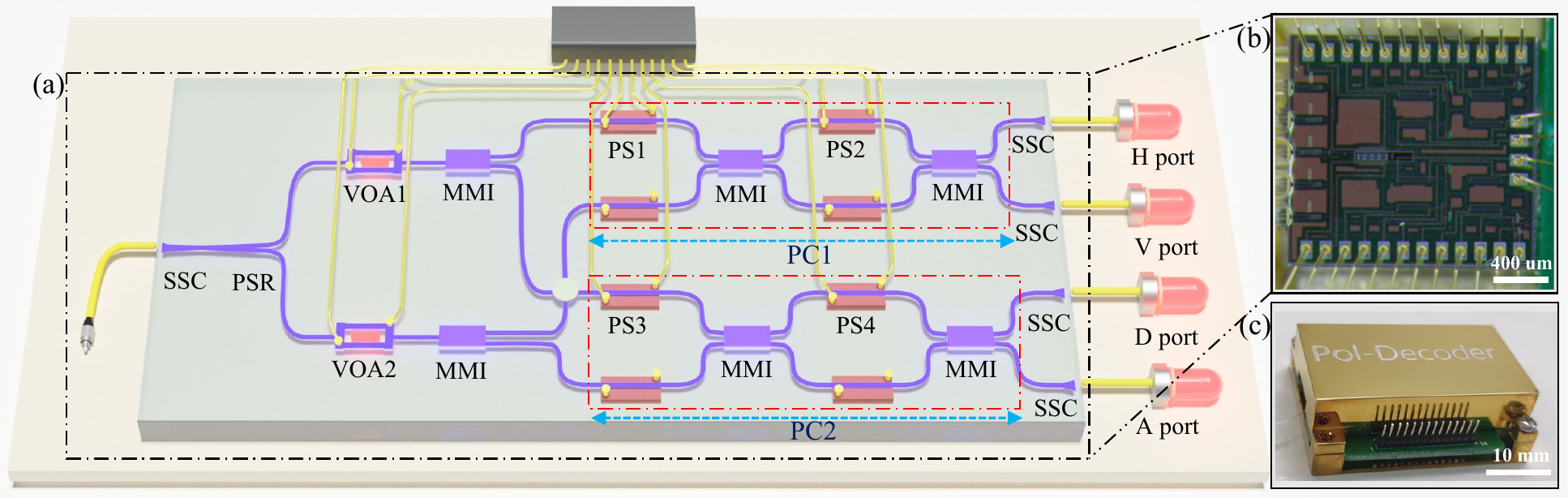}
		\caption{(a) Schematic of a silicon-based decoder chip. All devices are manufactured on standard silicon-based optoelectronic platforms, including the polarization splitter-rotator (PSR), thermal phase shifter (PS), multimode interferometer (MMI), and variable optical attenuation (VOA). The device contains two polarization controllers (PC1 and PC2). Photons moved through ports H, V, D, and A in $\left | H  \right \rangle$, $\left | V  \right \rangle$, $\left | D\right \rangle$, and $\left | A  \right \rangle$ states, respectively. (b) The microscopic image of the decoder chip. The size of the decoder chip is 1.6$\times$1.7 mm$^{2}$. (c) The physical picture of the decoder chip. The volume of the package is 3.95 $\times$ 2.19 $\times$ 0.90 cm$^{3}$.}
		\label{Chip_micrograph}
	\end{figure*}

	To apply the increasingly mature QKD technology to standard communication networks, it is important to develop a stable, simple, inexpensive, and miniaturized QKD system. Silicon photonics, the leading platform of quantum photonics technology, has excellent characteristics, such as high integration, mature technology, and compatibility with CMOS~\cite{Siew2021,Baets2021}. In recent years, silicon-based optoelectronic platforms have been used to develop high-speed, robust, and practical QKD devices. These integrated devices can implement the BB84 protocol~\cite{Ma2016,Sibson2017,Bunandar2018,Paraiso2021,Zhu2022,2022LOYS-chip,2022You-PLC}, measurement device--independent
	QKD protocol~\cite{Wei2020,Cao2020,Semenenko2020,2021Zheng-Chip}, continuous variable  QKD protocol~\cite{ZhangG2019}, high-dimensional  QKD protocol~\cite{Ding2017HD-QKD}, and coherent one-way  and differential phase shift  protocols ~\cite{Sibson2017,De2021}. The potential realistic vulnerabilities of chip-based devices have been extensively studied in previous studies~\cite{2018Li-PDL,2022Ye-chip,2021Tan-chip,2022Huang-PDL}. A recently published review provides a detailed report on the development of integrated QKD~\cite{Liu2022}.
	
	Polarization encoding has been extensively applied in QKD systems over fiber-based or free-space channels. Polarization encoding--based implementations of QKD have been extensively studied using bulk optical components~\cite{LiYang2019,Ma2021} or silicon photonics~\cite{Ma2016,Sibson2017-2,Bunandar2018,LinYang2022,Zhu2022}. However, because of the difficulty of realizing polarization conversion using silicon photonics, prior chip-based devices utilized either bulk optical devices to decode polarization states~\cite{Ma2016,Sibson2017-2,Bunandar2018,Zhu2022} or additional off-chip devices to perform polarization-based tracking or polarization compensation~\cite{Zhu2022, LinYang2022}.  
	
	In this study, we developed a novel silicon-based decoder for polarization-encoding QKD. Based on the polarization-to-path conversion method~\cite{2016Wangjianwei}, the chip avoids the requirement of using polarization-independent devices and realizes a polarization state analyzer and polarization compensation for the BB84 protocol without requiring additional hardware. The chip was manufactured using a standard silicon-based photonic platform and exhibited robustness against environmental disturbances, owing to its compact design. In an experimental stability test, an average quantum bit error rate (QBER) of $0.56\pm 0.02\%$ was achieved through continuous operation for 10 h without any active feedback. We further tested the stability by inducing a random fiber polarization scrambler to emulate the polarization drift. We showed that the decoder can automatically compensate for polarization drift using a developed feedback algorithm. Finally, we performed a proof-of-principle QKD demonstration with a finite-key secret rate of 240 bits per second (bps) over a 100-km fiber spool. Our study demonstrates the feasibility of a silicon-based integrated decoder chip and represents an important step toward a fully integrated polarization-encoding QKD system.
	
	The remainder of this paper is organized as follows: In Sec.~\ref{Silicon photon polarization decoder}, we describe the basic setup of the decoder chip. The results of the experiment are described in Sec.~\ref{experimental test}. Finally, we summarize the concluding points in Sec.~\ref{Conclusion}.

\section{Silicon photon polarization decoder}\label{Silicon photon polarization decoder}

A schematic of the proposed decoder chip is shown in Fig.~\ref{Chip_micrograph}(a). The chip was fabricated using a standard silicon photonics foundry with a size of 1.6$\times$1.7 mm$^{2}$, as shown in Fig.~\ref{Chip_micrograph}(b). It was packaged based on a chip-on-board assembly, with a total volume of 3.95 $\times$ 2.19 $\times$ 0.90 cm$^{3}$, as shown in Fig.~\ref{Chip_micrograph}(c). The entire decoder chip is composed of a spot-size converter (SSC), a polarization splitter-rotator (PSR), two variable optical attenuations (VOAs), and four Mach--Zehnder interferences (MZIs), each containing two multi-mode interferences (MMIs) and two thermal phase shifters (PSs). 

A spot-size converter (SSC) was used to couple the received randomly polarized signal light into the decoder chip. Subsequently, the PSR converted the horizontal and vertical components of the polarized signal light into two on-chip single-mode waveguides propagated as two transverse electric (TE) signals.
The variable optical attenuation (VOA) was placed into each arm of the PSR to balance the polarization-dependent loss caused by state preparation~\cite{2018Li-PDL}.

\begin{figure*} [!ht]
	\centering
	\includegraphics[scale=0.7]{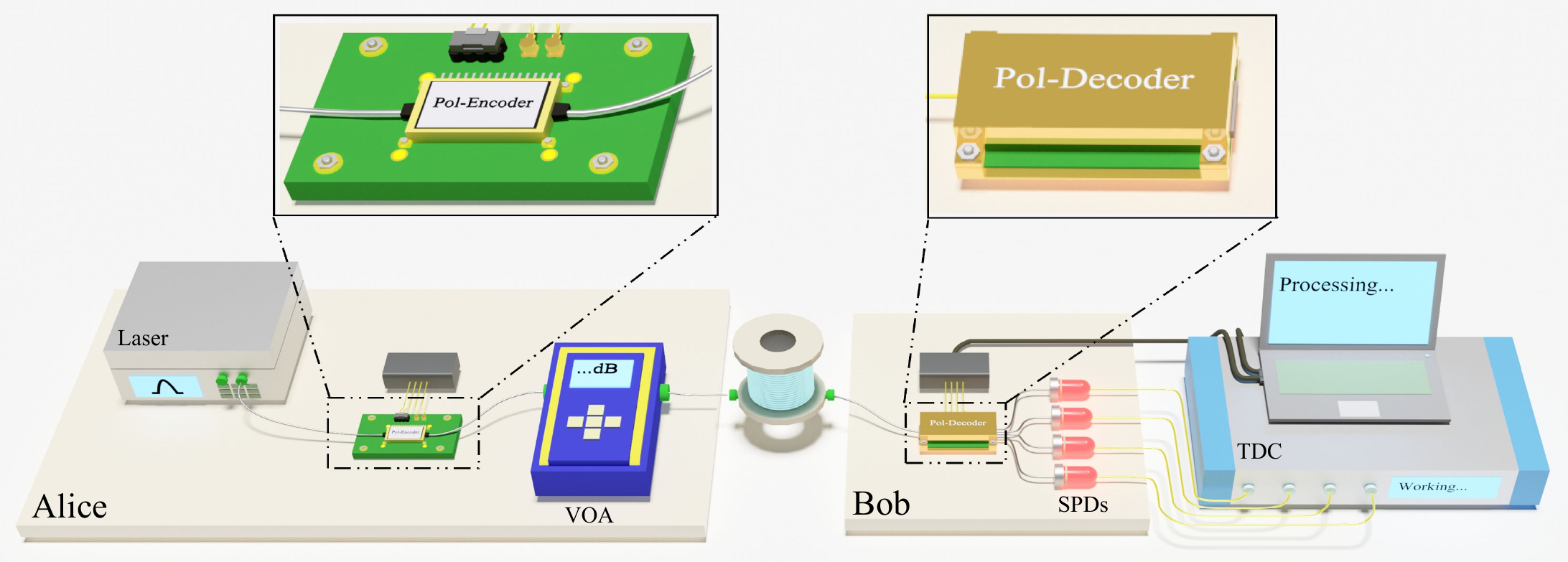}
	\caption{Experimental setup of the silicon-based integrated QKD. Alice utilizes a laser to generate pulsed light, with a repetition rate of 50 MHz. The generated pulsed light is coupled into an encoder chip containing an intensity modulator and a polarization modulator for the random modulation of decoy and polarization states. It is then attenuated to a single photon level by the off-chip VOA and transmitted to Bob via a fiber spool. Bob decodes the received photons using a decoder chip and detects them using four single-photon detectors (SPDs), and the detector results are recorded by the time-to-digital converter (TDC). The detector events recorded by the TDC are processed using a personal computer. The personal computer and the electronics module of the decoder chip are connected as a polarization feedback loop.}
	\label{SYSTEM}
\end{figure*}  

Each TE signal was then spit by a 1$\times$2 multi-mode interferences (MMIs) into either a stage 1 polarization controller (PC1) or stage 2 PC (PC2). Each PC comprised a phase shifter (PS)--driven Mach--Zehnder interference (MZI) with two additional PSs. The outputs of the PC were coupled with single-photon detectors via the SSC. Further information regarding the chip design and fabrication is provided in Appendix~\ref{appendix-chip-design}. By carefully controlling the PSs, the outputs of PC1 (PC2) constitute the positive operator value measurements (POVMs) in the Z-basis (X-basis) in the BB84 protocol. Furthermore, because the POVMs were constrained by actively adjusting the PS, this allows our chip to compensate for variations in the polarization states caused by fiber birefringence. 
Further information regarding the POVMs by the polarization decoder chip for the BB84 protocol is presented in  Appendix~\ref{appendix-POVM}.

  \section{ experimental test}\label{experimental test}
  
  To test the performance of the chip, we built a testing platform, as shown in Fig.~\ref{SYSTEM}. We used a commercial laser source (LD, WT-LD, Qasky Co. LTD), with a repetition rate of 50 MHz and a pulse width of 200 ps. The generated light pulses were coupled into the encoder chip, which was composed of an intensity modulator and a polarization modulator, generating decoy states with a dramatic extinction ratio (ER) of $\sim18$ dB and four polarization states with an average ER of $\sim25$ dB. The modulated quantum states were then attenuated to the single-photon level through off-chip variable optical attenuation (DA-100, OZ Optics Ltd.) and transmitted to the receiver chip over a fiber channel. The signals were analyzed by using the decoder chip and then detected by an off-chip single-photon detector (SPDs, WT-SPD2000, Qasky Co. LTD.), with a detection efficiency of $10 \%$ and a dark count rate of 400 Hz. Final detection events were recorded using a time-to-digital converter (TDC; quTAG100, GmbH). A personal computer was used to process the data recorded by the TDC.

  \subsection{Characterization of decoder chip }

\begin{figure}[ht!]
  	\centering
  	\includegraphics[width=0.9\linewidth]{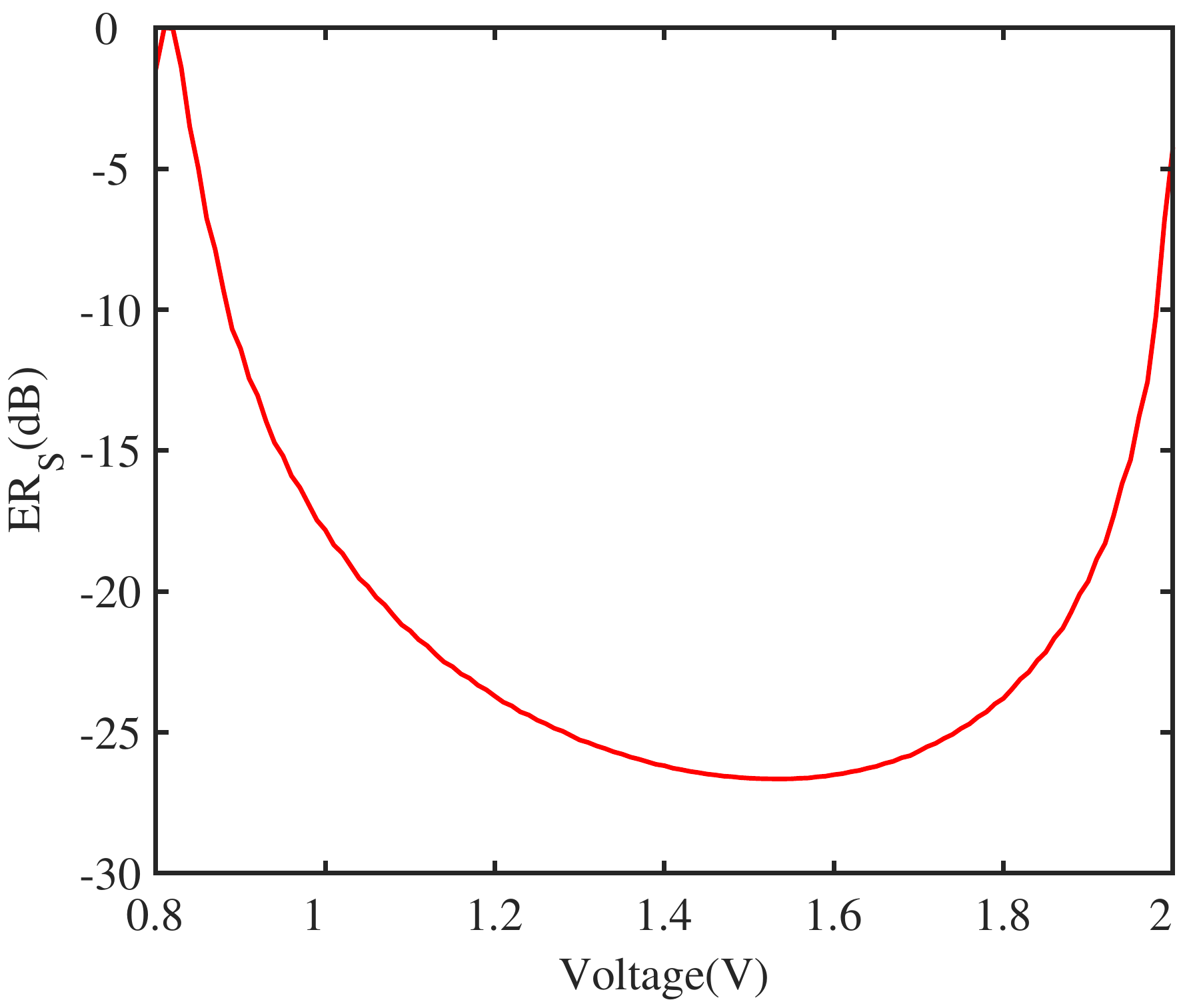}
  	\caption{Extinction ratios obtained by separately sweeping the PS voltages inside the intensity modulator. The abscissa represents the scanned voltage value, the ordinate is $ER_{S}=10\times lg({I_{out}/I_{min}})$, $I_{out}$ is the output photons number corresponding to the current voltage, and $I_{min}$ is the minimum output photons number in the entire scanning voltage range.}\label{ER}
  \end{figure}

 First, the decoder chip was characterized. The insertion loss was approximately 4.6 dB, which is comparable to that reported in previous studies~\cite{ZhouXiaoqi2020, LinYang2022}. The 3-dB bandwidth of PSs was approximately 3 kHz, which was sufficient to respond to the polarization variations in field-buried
  and aerial fibers~\cite{2017Ding-polarization}.
  We then measured the half-wave voltage of the PS, which is a key parameter of the chip. Here, we measured the half-wave voltage of the PS by measuring the ER of an MZI containing PS (using the same technology as the decoder chip). As shown in Fig.~\ref{ER}, The IM has a maximum static ER of approximately 26.65 dB and a half-wave voltage PS of approximately 0.72 V.
  
  Finally, we tested the ERs of the decoder chip to measure different polarization states. The pulsed light emitted by the laser was coupled into the encoder chip to generate 
  one of the four BB84 polarization states and then transmitted to the decoder chip for demodulation. ERs of 28.86 dB, 26.02 dB, 27.70 dB, and 29.21 dB were obtained for the polarization states $\left | H  \right \rangle ,\left | V \right \rangle ,\left | D \right \rangle$ and $\left | A  \right \rangle$, respectively.

 \subsection{Inherent stability test }

\begin{figure}[h]
  	\centering
  	\includegraphics[width=0.9\linewidth]{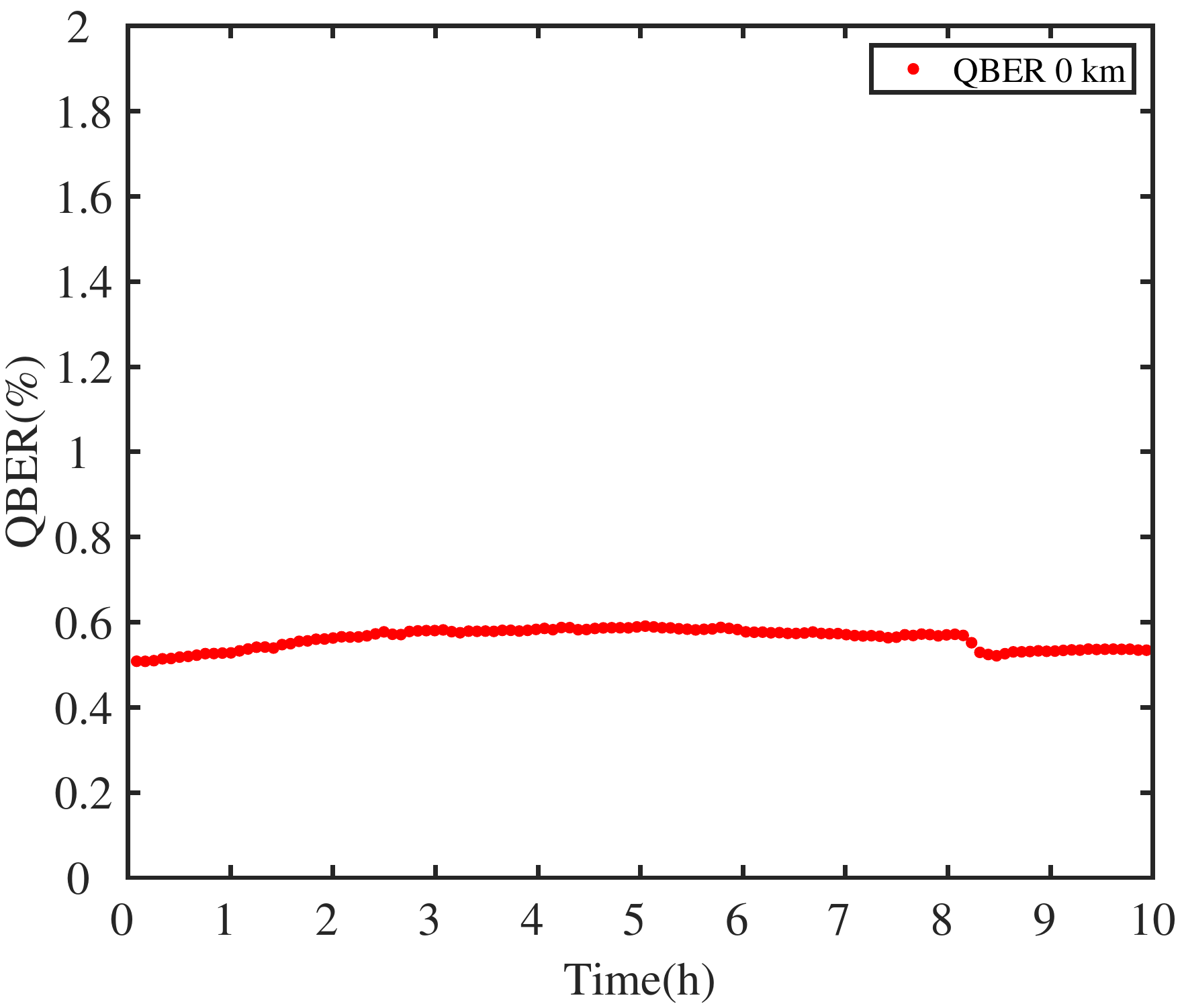}
  	\caption{QBER of the polarization-encoding decoder without active polarization feedback over 10 h. The total QBERs are represented by red lines. QBER is calculated every 5 min.}\label{Stability}
  \end{figure}

 To demonstrate the reliability of the decoder chip, a 10-h long-term run test was conducted. We randomly generated four BB84 polarization states and measured them directly using the decoder chip, as shown in Fig.~\ref{Stability}. An average QBER of $0.56\pm 0.02\%$ was obtained through continuous operation for 10 h without any active feedback control. Our chip exhibited inherent stability and provided the feasibility of building a QKD system with a long run time.
  
  \subsection{Automatic polarization compensation over a 75-km fiber link }
  
  \begin{figure}[t]
  	\centering
  	\includegraphics[width=0.9\linewidth]{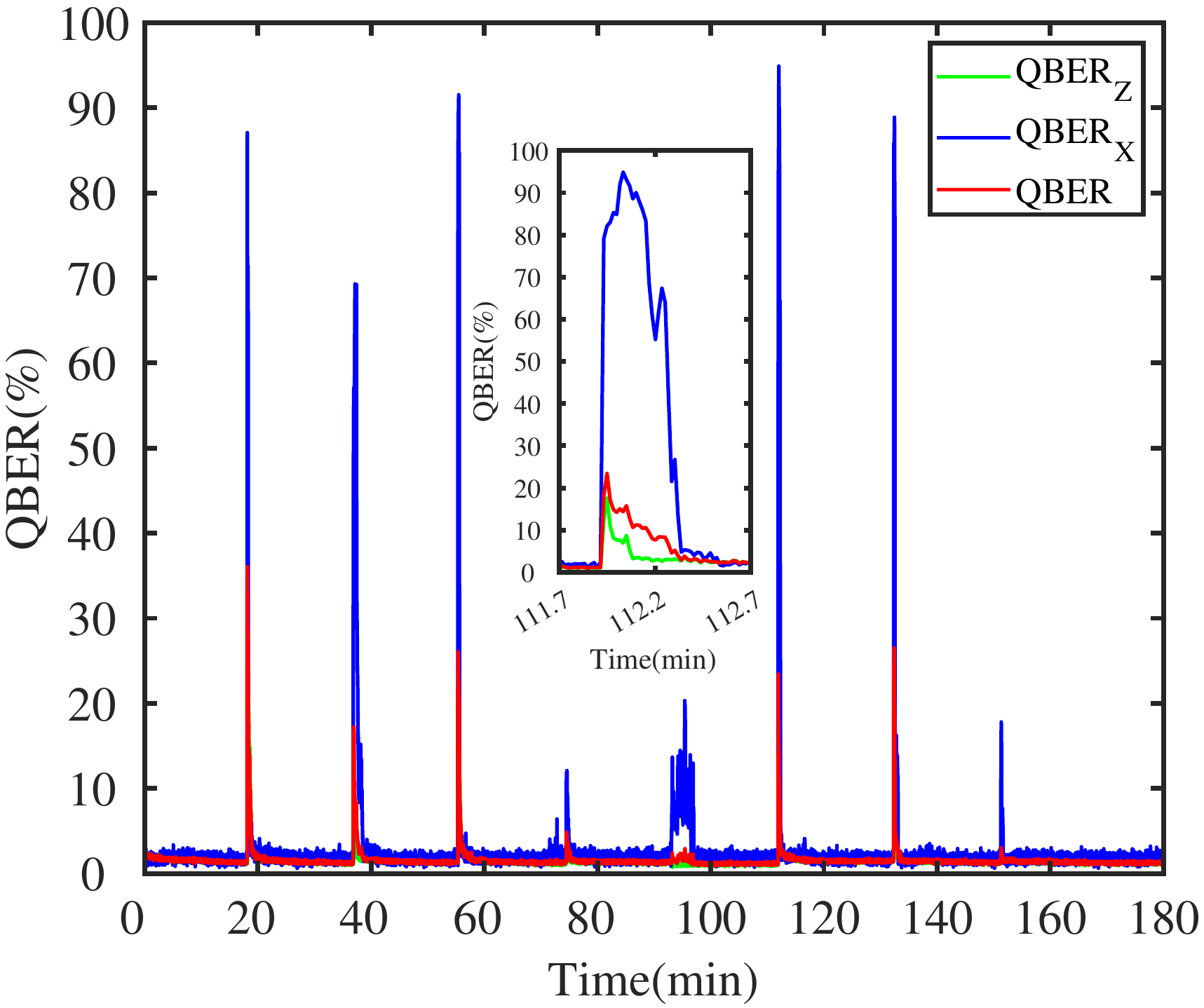}
  	\caption{Polarization compensation system when the polarization scrambler performs a random disturbance at any time within 20--30 min. The red solid line represents the average QBER of the system, whereas the green and blue lines represent the QBER $_{Z}$ in the Z-basis and QBER $_{X}$ in the X-basis, respectively. The inset shows enlarged experimental data during the period of 111.7--112.7 min.
  	}\label{PFB_disturbance}
  \end{figure}

 We developed a polarization compensation scheme that can realize real-time polarization compensation using only shared qubits. The basic idea of the polarization compensation scheme is that Bob evaluates the average QBER  Z- and X-bases (QBER $_{Z}$ and QBER $_{X}$) every second. The computer, after processing by the feedback algorithm (the algorithm is based on gradient descent~\cite{MaMinglei2020}), then controls the programmable linear DC source to change the retardance of $\theta_{1}$, $\theta_{2}$, $\theta_{3}$, and $\theta_{4}$ until QBER $_{Z}$ and QBER$_{X}$ are less than the set threshold. The detailed process of our scheme is described in Appendix~\ref{appendix-GD}.

  To test the performance of our polarization compensation scheme, we randomly modulated four BB84 polarization states and used the VOA to attenuate to approximately 0.6 photons per pulse, which were then sent to Bob via a 75-km fiber spool for decoding and detection. Because the polarization drift in an installed fiber is relatively small, a polarization scrambler was placed behind the fiber spool to randomly disturb the polarization of photons entering the decoder chip (the scrambler triggers the time interval randomly within 20--30 min).
  
  The experimental results are presented in Fig.~\ref{PFB_disturbance}. QBER $_{Z}$, QBER $_{X}$, and the total QBER of the system are represented by green, blue, and red lines, respectively. During the entire 180-min running time, the error rate increased by more than $50\%$ on average after each perturbation. The polarization compensation system feedback reduced the average QBER to approximately $1.39\%$ in about 1 min, as shown in the inset of Fig.~\ref{PFB_disturbance}. The polarization can be compensated potentially  in much shorter time by optimizing algorithm. The results showed that the polarization compensation scheme is both feasible and robust.
  
  \subsection{QKD Secure key rates at different fiber distances}

 Finally, we performed a QKD test using fiber distances of 25 km, 50 km, 75 km, and 100 km. For each distance, we performed a real one decoy--state protocol~\cite{Hugo2018} with optimized parameters, except for the probability of Bob choosing the Z-basis, and the X-basis was set to 0.5 (owing to the balanced MMI used in our chip). 
  
  For each distance, the total number of pulses $N=10^{10} $ and the SKR in the finite regime were estimated by, 
  \begin{equation}\label{R_finite}
  	\begin{split}
  		l \leq s_{z, 0}^{l}& +s_{z, 1}^{l}\left(1-h\left(\phi_{z}^{u}\right)\right)\\&-\lambda_{\mathrm{EC}} -6 \log _{2}\left(19 / \epsilon_{\mathrm{sec}}\right)-\log _{2}\left(2 / \epsilon_{\mathrm{cor}}\right) ,           	
  	\end{split}  
  \end{equation}
  where $s_{z,0}^{l} $ is the lower bound of the detector event received by Bob, given that Alice sends an empty state under the Z-basis, $s_{z,1}^{l} $ is Bob receiving the lower bound of the detection event, given that Alice only sends a single-photon state in the Z-basis. $\phi _{z}^{u} $ is the upper bound of the phase error rate, $\lambda  _{EC}$ is the number of bits disclosed for error correction, and $\epsilon_{sec}$ and $\epsilon_{cor}$ are the parameters used to evaluate secrecy and correctness, respectively. The $h\left ( x \right )=-x\log_{2}{x}-\left ( 1-x \right )\log_{2}{\left ( 1-x \right ) }$ denotes the binary Shannon information function.

\begin{figure}[h]
  	\centering
  	\includegraphics[width=0.9\linewidth]{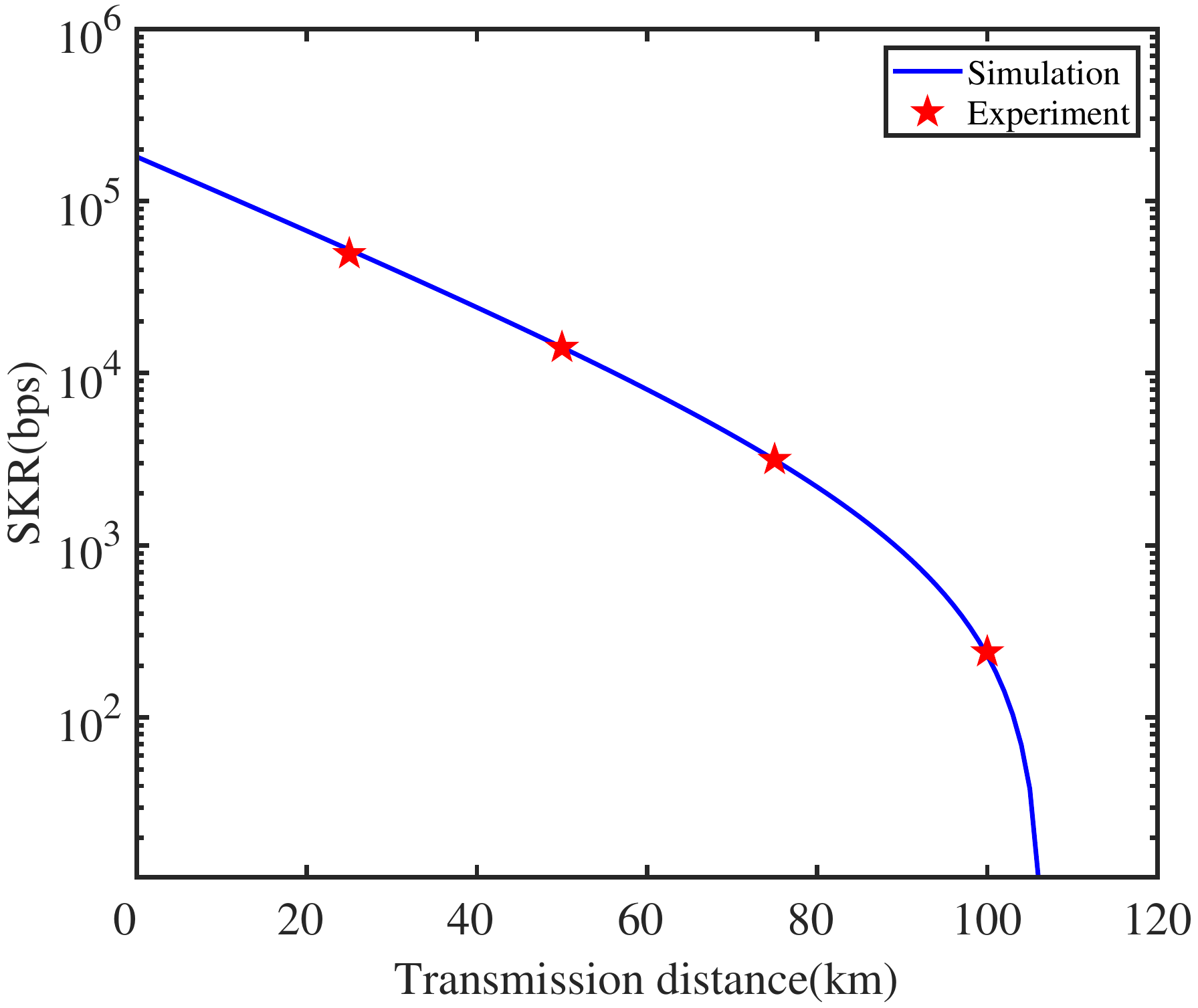}
  	\caption{Secure key rate versa fiber distance. The red pentagons represent the experimental results, and the blue lines represent the theoretical simulation results based on our experimental parameters.}\label{QKD}
  \end{figure}

  The experimental results are plotted in Fig.~\ref{QKD}. We achieved an SKR of 240 bps over 100 km. The further detailed experimental results for each distance are shown in Appendix~\ref{appendix-DECOY}.

  \section{Conclusion}\label{Conclusion}
  
 In this study, we developed and validated a novel decoder chip for polarization-encoding QKD systems using a silicon-based optoelectronic platform. The receiver chip shows high inherent stability and can demodulate the polarization states with favorable extinction ratios. The chip can automatically perform polarization compensation. In the QKD test, secure key bits were successfully distributed over a fiber distance of up to 100 km. 
 
 These results demonstrate the feasibility of implementing polarization decoder chips without an off-chip polarization controller, which is an important step toward a more compact QKD system. Our chip would be attractive across different operative scenarios, in particular, an uplinking satellite-based QKD, where the receiver is placed in space~\cite{2021Uplinking}.

\section{Acknowledgments}
 We thank Shizhuo Li for drawing the diagram of the chip. This study was supported by the National Natural Science Foundation of China (Nos. 62171144, 62031024, and 11865004), the Guangxi Science Foundation (No.2021GXNSFAA220011), and the Open Fund of IPOC (BUPT) (No. IPOC2021A02).

  \appendix 
  
  \section{Device design and fabrication}\label{appendix-chip-design}
  
  The decoded chip had a device footprint of 1.6$\times$1.7 mm$^2$. The device was fabricated on a standard silicon-on-insulator wafer with a 220-nm silicon layer and a 3-\textmu m buried silica oxide layer. The width of the single-mode silicon waveguide was 450 nm. The light was coupled into/out of the device via a spot-size converter (SSC) with a taper length of approximately 100 \textmu m and then split and converted into transverse electric (TE) modes using a compact polarization rotator-splitter~\cite{2021Chen}. The variable optical attenuations (VOAs) function based on the forward carrier injection PIN junction, the attenuation of which increases with the applied voltage. The length of the VOA was approximately 200 \textmu m. The 1$\times$2 multimode interferometer (MMI) and 2$\times$2 MMI couplers were designed to be approximately 4$\times$14 \textmu m$^2$ and 8$\times$60 \textmu m$^2$, respectively, to obtain a nearly balanced splitting ratio. All eight phase shifters (PSs) shown in Fig.~\ref{Chip_micrograph}(a) are identical, with a length of 260 \textmu m, which efficiently results in a static extinction of approximately 28 dB when implemented in Mach--Zehnder interferences (MZIs). Notably, in real-time, only one PS on one arm in each MZI was active and the other PS was designed for the compensation of a 0.05-dB loss. The pitch of the aluminum DC pad was approximately 150 \textmu m. The pad area was approximately 80$\times$100 \textmu m$^2$.

 \section{POVM of the decoder chip}\label{appendix-POVM}
 
Here, we describe the construction of positive operator value measurements (POVMs) for the BB84 protocol by controlling the PSs in the chip. For simplification, we did not consider the fiber’s polarization drift, that is, four ideal BB84 polarization states, $\left | H  \right \rangle$, $\left | V  \right \rangle$, $\left | D  \right \rangle$ and $\left | A  \right \rangle$ , are transmitted to the decoder chip. Arbitrarily polarized light incident on the decoder chip was first converted by the PSR from the polarization-encoding state to the path-encoding state, and this process is expressed as,

\begin{equation}\label{path}
	\begin{split}
		|\phi\rangle_{\mathrm{pol}}=\alpha|H\rangle+\beta|V\rangle \Rightarrow|\varphi\rangle_{path}=\left(\begin{array}{l}
			\alpha \\
			\beta
		\end{array}\right).
	\end{split}
\end{equation}
When we consider the ideal BB84 states, $ (\alpha, \beta) $ is  $(1,0) $ for  $|H\rangle,~ (0,1)$  for  $|V\rangle,~(\sqrt{2} / 2, \sqrt{2} / 2) $ for $|D\rangle $ and  $(\sqrt{2} / 2, -\sqrt{2} / 2)$  for  $|A\rangle$. The two $ 1 \times 2$ MMI extend the two-path state to a four-path state and then guide two of them into PC1 and the other two into PC2. The corresponding splitter operators are given by,
\begin{equation}\label{SZX}
	\begin{split}
		S  = & \frac{\sqrt{2}}{2} \left(\begin{array}{cc}
			1 & 0 \\
			0 & 1 \\
		\end{array}\right).
	\end{split}
\end{equation}
The operators of PS and $ 2 \times 2$ MMI can be described by the Jones matrix as,

\begin{equation} \label{ARU}
	\begin{split}
		R_{\theta}=\left(\begin{array}{cccc}
			e^{i \theta} & 0 \\
			0 & 1
		\end{array}\right) 
		,~             	   U=\frac{\sqrt{2}}{2}\left(\begin{array}{cccc}
			1 & i  \\
			i & 1
		\end{array}\right).
	\end{split}
\end{equation}
The collapse results of ports H and V on the Z-basis are
$\left|\psi_{H}\right\rangle=\left(\begin{array}{llll}1 & 0 \end{array}\right)^{\mathrm{T}}, \left|\psi_{V}\right\rangle=\left(\begin{array}{llll}0 & 1\end{array}\right)^{\mathrm{T}}$, respectively. The collapse results of ports D and A in the X-basis are $\left|\psi_{D}\right\rangle=\left(\begin{array}{llll} 1 & 0\end{array}\right)^{\mathrm{T}} ,    \left|\psi_{A}\right\rangle=\left(\begin{array}{llll} 0 & 1\end{array}\right)^{\mathrm{T}} $, respectively, where  $\mathrm{T} $ is the transposition symbol.

\begin{table}[]
	\centering
	\caption{The measurement probability of different polarization states, where  $\left | H  \right \rangle ,\left | V  \right \rangle ,\left | D  \right \rangle$, and $\left | A  \right \rangle, $ respectively, represent the polarization state of the photons sent by Alice. The $P_{H}, P_{V} $, $P_{D}$, and $P_{A} $ represent the measured probabilities at ports H, V, D, and A, respectively.}
	\renewcommand\arraystretch{1.2}
	\tabcolsep=0.4cm
	\scalebox{1}{\begin{tabular}{cccccc}
			\hline\hline
			\multirow{2}{*}{Input state } & &      \multicolumn{4}{c}{Measurement probability} 
			\\  \cline{3-6} 
			&  &~$P_{H}$~&~$P_{V}$~&~$P_{D}$~& ~$P_{A}$~   
			\\ \hline 
			$\left | H  \right \rangle$   &  & 0.5 & 0 &0.25 & 0.25 
			\\
			$\left | V \right \rangle$     &   & 0 & 0.5 & 0.25 & 0.25 
			\\
			$\left | D  \right \rangle$    &  & 0.25 & 0.25 & 0.5 & 0
			\\
			$\left | A  \right \rangle$   &   & 0.25 & 0.25 & 0 & 0.5
			\\ \hline\hline
		\end{tabular}
		\label{measurement-probability}
	}
\end{table}

Assuming that the retardance of PS1, PS2, PS3, and PS4 in Fig.~\ref{Chip_micrograph} are $\theta_{1}, \theta_{2}, \theta_{3}$, and $\theta_{4}$, respectively, the measurement operator corresponding to ports H, V, D, and A is,
\begin{equation}\label{A-MZH}
	\begin{split}	
		M_{H}&=\sqrt{S^{*} R_{\theta _{1} }^{*} U^{*}R_{\theta _{2}}^{*} U^{*} 
			\left|\psi_{H}\right\rangle\left\langle\psi_{H}\right| 
			UR_{\theta _{2} } UR_{\theta _{1} }S} \\&=\frac{\sqrt{2}}{4}\left(\begin{array}{cccc}
			~1-\cos \theta _{2}~ & e^{-i\theta _{1} } \sin \theta _{2}~   \\
			~e^{i\theta _{1} } \sin \theta _{2}~  & 1+\cos \theta _{2}~
		\end{array}\right),
	\end{split}
\end{equation}

\begin{equation}\label{A-MZV}
	\begin{split}	
		M_{V}&=\sqrt{S^{*} R_{\theta _{1} }^{*} U^{*}R_{\theta _{2}}^{*} U^{*} 
			\left|\psi_{V}\right\rangle\left\langle\psi_{V}\right| 
			UR_{\theta _{2} } UR_{\theta _{1} }S}\\&=\frac{\sqrt{2}}{4}\left(\begin{array}{cccc}
			1+\cos \theta _{2} & -e^{-i\theta _{1} } \sin \theta _{2}   \\
			-e^{i\theta _{1} } \sin \theta _{2}  & 1-\cos \theta _{2}
		\end{array}\right),
	\end{split}
\end{equation}

\begin{equation}\label{A-MXD}
	\begin{split}	
		M_{D}&=\sqrt{S^{*} R_{\theta _{3} }^{*} U^{*}R_{\theta _{4}}^{*} U^{*} 
			\left|\psi_{D}\right\rangle\left\langle\psi_{D}\right| 
			UR_{\theta _{4} } UR_{\theta _{3} }S}    \\&=\frac{\sqrt{2}}{4}\left(\begin{array}{cccc}
			~1-\cos \theta _{4}~ & e^{-i\theta _{3} } \sin \theta _{4}~   \\
			~e^{i\theta _{3} } \sin \theta _{4}~  & 1+\cos \theta _{4}~
		\end{array}\right),
	\end{split}
\end{equation}

\begin{equation}\label{A-MXA}
	\begin{split}	
		M_{A}&=\sqrt{S^{*} R_{\theta _{3} }^{*} U^{*}R_{\theta _{4}}^{*} U^{*} 
			\left|\psi_{A}\right\rangle\left\langle\psi_{A}\right| 
			UR_{\theta _{4} } UR_{\theta _{3} }S}\\&=\frac{\sqrt{2}}{4}\left(\begin{array}{cccc}
			1+\cos \theta _{4} & -e^{-i\theta _{3} } \sin \theta _{4}   \\
			-e^{i\theta _{3} } \sin \theta _{4}  & 1-\cos \theta _{4}
		\end{array}\right).
	\end{split}
\end{equation}
These positive semi-definite operators satisfy the following completeness relation:

\begin{equation}\label{ID}
	\begin{split}	
		M_{H}^{*} M_{H}+M_{V}^{*} M_{V}+M_{D}^{*} M_{D}+M_{A}^{*} M_{A}=I,
	\end{split}
\end{equation}
where $I$ is the identity matrix and $*$ is the conjugate transpose notation.

Therefore, for any polarization state, the measurement probability at ports H, V, D, and A is given by,

\begin{table*}[ht!]
	\centering
	\caption{Symbol definitions of the polarization feedback algorithm}
	\renewcommand\arraystretch{1.3}
	\tabcolsep=0.1cm
	\scalebox{1}
	{\begin{tabular}{l}
			\hline \hline 
			$Definitions:$ \\
			$V_{i}(PS_{j} ) :$ The voltages loaded on $PS_{j}$ in the $i$th cycle, $ j \in \{1, 2, 3, 4\}$.\\
			$E_{\lambda,i}:$ The QBER of the $\lambda$ basis in the $i$th cycle, $\lambda \in \{Z,X\} $.\\ 
			$G_{j, i}^{\lambda}:$ The partial derivative  is used to estimate the change of $E_{\lambda,i}$, where $i$ is the number\\~~~~~~~~~ of program cycles, $ j \in \{1, 2, 3, 4\}$.\\
			$\Delta v_{i} :$ A tiny dither voltage is applied to tune $V_{i}(PS_{j} )$.  $V_{i}(PS_{j} )$ is the voltage at the $j$th\\~~~~~~~~ PS in the $i$ cycle, $ j \in \{1, 2, 3, 4\}$.\\
			$\Delta \alpha_{i}:$ A scalar factor is used to tune $V_{i}(PS_{j} )$. $V_{i}(PS_{j} )$ is the voltage at the $j$th \\~~~~~~~~~PS in the $i$ cycle, $ j \in \{1, 2, 3, 4\}$.\\
			$E_{\lambda,th}:$ The threshold of QBER in the $\lambda$ basis,  $\lambda \in \{Z,X\} $.\\
			\hline \hline
	\end{tabular}}
	\label{symbol_definitions}
\end{table*}

\begin{equation}\label{P-path-H}
	\begin{split}	
		P_{H}\left(\left|\varphi_{p a t h}\right\rangle\right)&=\left\langle\varphi_{p a t h}\left|M_{H }^{*} M_{H }\right| \varphi_{p a t h}\right\rangle\\&=\frac{1}{4}\left [ \left ( 1-\cos \theta _{2}  \right ) \alpha ^{*} \alpha + e^{i\theta _{1}}\sin \theta _{2}\alpha \beta ^{*}\right.\\&\phantom{=\;\;}\left. + e^{-i\theta _{1}}\sin \theta _{2} \alpha ^{*} \beta +\left ( 1+\cos \theta _{2}  \right ) \beta^{*} \beta  \right ],   
	\end{split}
\end{equation}

\begin{equation}\label{P-path-V}
	\begin{split}	
		P_{V}\left(\left|\varphi_{p a t h}\right\rangle\right)&=\left\langle\varphi_{p a t h}\left|M_{V }^{*} M_{V }\right| \varphi_{p a t h}\right\rangle\\&=\frac{1}{4}\left [ \left ( 1+\cos \theta _{2}  \right ) \alpha ^{*} \alpha - e^{i\theta _{1}}\sin \theta _{2}\alpha \beta ^{*}\right.\\&\phantom{=\;\;}\left. -e^{-i\theta _{1}}\sin \theta _{2} \alpha ^{*} \beta +\left ( 1-\cos \theta _{2}  \right ) \beta^{*} \beta  \right ],  
	\end{split}
\end{equation}

\begin{equation}\label{P-path-D}
	\begin{split}	
		P_{D}\left(\left|\varphi_{p a t h}\right\rangle\right)&=\left\langle\varphi_{p a t h}\left|M_{D}^{*} M_{D }\right| \varphi_{p a t h}\right\rangle\\&=\frac{1}{4}\left [ \left ( 1-\cos \theta _{4}  \right ) \alpha ^{*} \alpha + e^{i\theta _{3}}\sin \theta _{4}\alpha \beta ^{*}\right.\\&\phantom{=\;\;}\left.+ e^{-i\theta _{3}}\sin \theta _{4} \alpha ^{*} \beta +\left ( 1+\cos \theta _{4}  \right ) \beta^{*} \beta  \right ],  
	\end{split}
\end{equation}

\begin{equation}\label{P-path-A}
	\begin{split}	
		P_{A}\left(\left|\varphi_{p a t h}\right\rangle\right)&=\left\langle\varphi_{p a t h}\left|M_{A }^{*} M_{A }\right| \varphi_{p a t h}\right\rangle\\&=\frac{1}{4}\left [ \left ( 1+\cos \theta _{4}  \right ) \alpha ^{*} \alpha - e^{i\theta _{3}}\sin \theta _{4}\alpha \beta ^{*} \right.\\&\phantom{=\;\;}\left. -e^{-i\theta _{3}}\sin \theta _{4}  \alpha ^{*} \beta +\left ( 1-\ cos \theta _{4}  \right ) \beta^{*} \beta  \right ].  
	\end{split}
\end{equation}

From Eqs~(\ref{P-path-H})-(\ref{P-path-A}), we find that, by carefully adjusting the phase of PSs where $\theta_{1}=0$, $\theta_{2}=\pi$, $\theta_{3}=0$, $\theta_{4}=\pi/2$, the decoder chip can constitute the POVM of the Z- and X-bases, and the measurement probabilities of different incident states at ports H, V, D, and A are shown in Table~\ref{measurement-probability}. 

These results indicated that our chip can compensate for variations in polarization states caused by fiber birefringence by adjusting the PS phases. Let us consider the POVM of the Z-bases as an example. Because of the polarization drift, the ideal BB84 states sent by Alice are in $\left | H  \right \rangle$, $\left | V  \right \rangle$, $\left | D  \right \rangle$ and $\left | A  \right \rangle$ degenerate into arbitrary polarization states before entering the decoder chip, which can be expressed as~\cite{Ding2017, 2018zhang-RSOP},

\begin{equation}\label{input}
	\begin{split} 
			\left | \varphi   \right \rangle_{path} ^{H}=	\begin{array}{l}
			\left[\begin{array}{cc}
				\cos \varphi  \\
				\sin \varphi  e^{i \phi}
			\end{array}\right] 
		\end{array},~~~~~~~~~~~~~~~~~\\ 
	\left | \varphi   \right \rangle_{path} ^{V}=	\begin{array}{l}
			\left[\begin{array}{cc}
				-\sin \varphi  e^{-i \phi}  \\
				\cos \varphi
			\end{array}\right] 
		\end{array}, ~~~~~~~~~~~~\\ 		\left | \varphi   \right \rangle_{path} ^{D}=\frac{\sqrt{2} }{2} 	\begin{array}{l}
			\left[\begin{array}{cc}
				\cos \varphi-\sin \varphi  e^{-i \phi}  \\
				\sin \varphi  e^{i \phi} + \cos \varphi
			\end{array}\right] 
		\end{array},
	\\
		\left | \varphi   \right \rangle_{path} ^{A}=\frac{\sqrt{2} }{2} 	\begin{array}{l}
			\left[\begin{array}{cc}
				\cos \varphi + \sin \varphi  e^{-i \phi}  \\
				\sin \varphi  e^{i \phi} - \cos \varphi
			\end{array}\right] 
		\end{array},
	\end{split}
\end{equation}
where $\varphi$ is the drift angle of the polarization basis between Alice and Bob and $\phi$ is the retardation between $\left | H  \right \rangle $ and $\left | V  \right \rangle $ components.
Using a similar analysis, we found that, when $\theta _{1} =\phi $, $\theta _{2} =\pi -2\varphi $, the arbitrary state $\left | \varphi   \right \rangle_{path} ^{H}$ $(\left | \varphi   \right \rangle_{path} ^{V})$ will be the output from port H with a probability of $50\%$ $(0)$ and from port V with a probability of $0 $ $(50\%)$. Furthermore, the arbitrary state $\left | \varphi   \right \rangle_{path} ^{D}$ $(\left | \varphi   \right \rangle_{path} ^{A})$ will be the output from port H with a probability of $25\%$ $(25\%)$ and from port V with a probability of $25\% $ $(25\%)$. This satisfies the Z-basis POVM requirement in the BB84 protocol.

\section{Polarization feedback algorithm}\label{appendix-GD}
For the Polarization-encoding QKD system, Alice and Bob share the same reference frame and need to compensate for the polarization drift caused by the fiber channel. In this study, we developed an automatic polarization compensation scheme based on the receiver chip setup, as shown in Fig.~\ref{PFB-flowchart}. Before implementing the polarization feedback scheme, Alice randomly sent four polarization states. The symbol definitions used in the polarization feedback algorithm are listed in Table~\ref{symbol_definitions}, which can be summarized as follows:

\begin{figure}[ht!]
	\centering
	\includegraphics[width=0.8\linewidth]{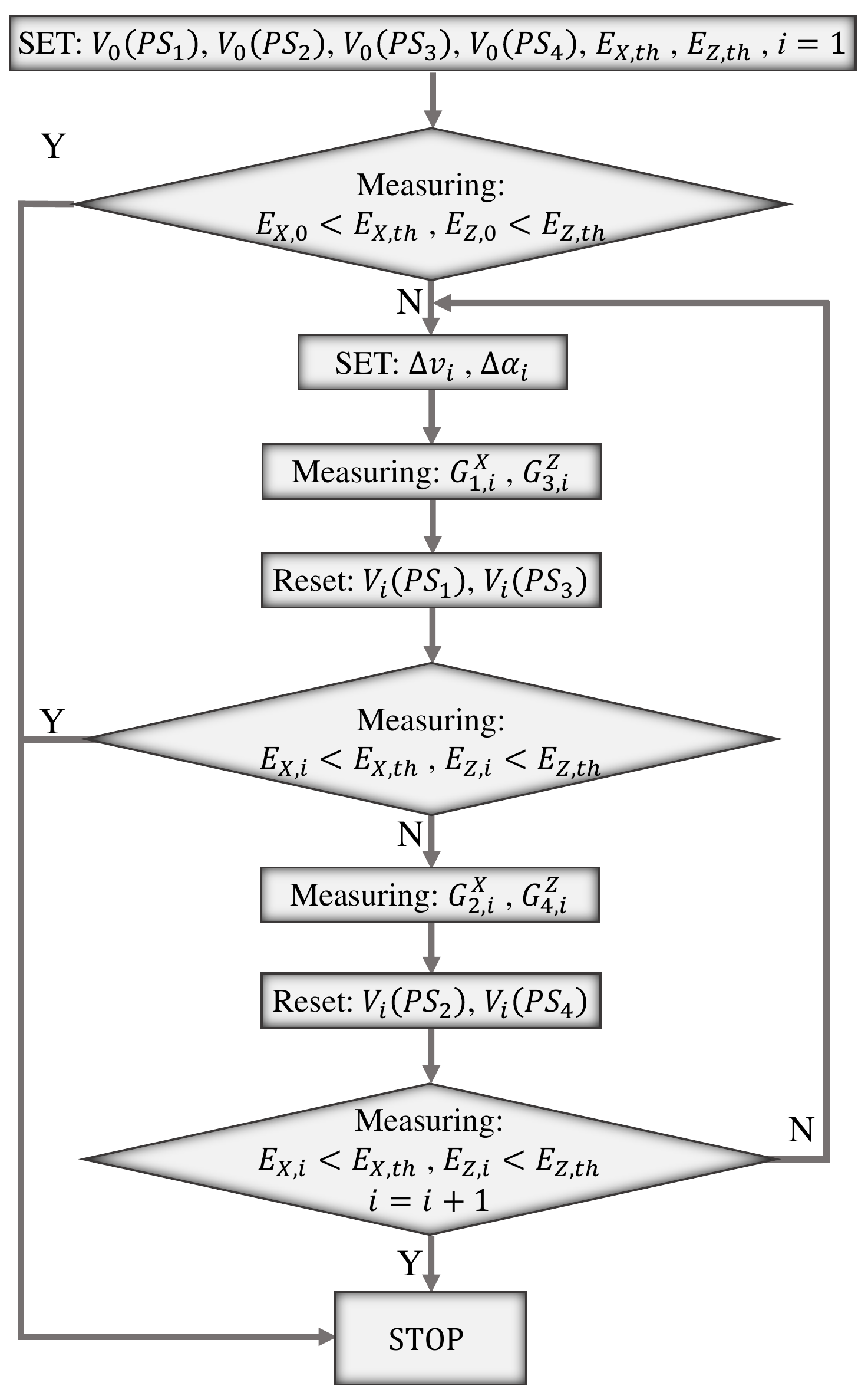}
	\caption{Flow chart showing the improved gradient descent method}\label{PFB-flowchart}
\end{figure}
Step 1: Set the values of $V_{0}(PS_{1})$, $V_{0}(PS_{2})$, $V_{0}(PS_{3})$, and $V_{0}(PS_{4})$ and set the thresholds of $E_{X}$ and $E_{Z}$ as $E_{X,th}$, and $E_{Z,th}$, respectively.

Step 2: Evaluate the systems $E_{X}$ and $E_{Z}$ and compare them with the set thresholds. If they are less than the set threshold, exit the feedback program; however, if they are greater than the set threshold, provide feedback.

Step 3: Set $\Delta v_{i} $ and $\Delta  \alpha_{i} $ based on the $E_{X,i}$ and $E_{Z,i}$ values. 

Step 4: Calculate the partial derivatives of $E_{X,i}$ and $E_{Z,i}$ as follows:

	\begin{equation} \label{GD13}
	\begin{split}
		G_{1, i}^{Z}=&\frac{E_{Z}\left(V_{i}(PS_{1})+\Delta v_{i}, V_i(PS_{2})\right)}{{2 \Delta v_{i}}} \\&- \frac{E_{Z}\left(V_{i}(PS_{1})-\Delta v_{i}, V_i(PS_{2})\right)}{{2 \Delta v_{i}}} \\
		G_{3, i}^{X}=&\frac{E_{X}\left(V_{i}(PS_{3})+\Delta v_{i}, V_i(PS_{4})\right)}{{2 \Delta v_{i}}} \\&- \frac{E_{X}\left(V_{i}(PS_{3})-\Delta v_{i}, V_i(PS_{4})\right)}{{2 \Delta v_{i}}}
	\end{split}
\end{equation}

Step 5: Apply the voltage to $PS_{1}$; $PS_{3}$ is given by,

\begin{equation}\label{PS13}
	\begin{split}
		V_{i}(PS_{1})=V_i(PS_{1})-\alpha_{i} G_{1, i}^{Z}, \\
		V_{i}(PS_{3})=V_i(PS_{3})-\alpha_{i} G_{1, i}^{X}.
	\end{split}
\end{equation}

Step 6: Evaluate the current $E_{X,i}$ and $E_{Z,i}$. If both values are below the set threshold, exit the feedback program; otherwise, continue with the following steps.

Step 7: Calculate the partial derivatives of $E_{X,i}$ and $E_{Z,i}$ as follows:

	\begin{equation}\label{GD24}
	\begin{split}
		G_{2, i}^{Z}=&\frac{E_{Z}\left(V_{i}(PS_{1}), V_i(PS_{2})+\Delta v_{i}\right)}{{2 \Delta v_{i}}} \\&- \frac{E_{Z}\left(V_{i}(PS_{1}), V_i(PS_{2})-\Delta v_{i}\right)}{{2 \Delta v_{i}}} \\
		G_{4, i}^{X}=&\frac{E_{X}\left(V_{i}(PS_{3}), V_i(PS_{4})+\Delta v_{i}\right)}{{2 \Delta v_{i}}} \\&- \frac{E_{X}\left(V_{i}(PS_{3}), V_i(PS_{4})-\Delta v_{i}\right)}{{2 \Delta v_{i}}}
	\end{split}
\end{equation}

Step 8: Apply the voltage to $PS_{2}$; $PS_{4}$ is given by,

\begin{equation}\label{PS24}
	\begin{split}
		V_{i}(PS_{2})=V_i(PS_{2})-\alpha_{i} G_{2, i}^{Z}, \\
		V_{i}(PS_{4})=V_i(PS_{4})-\alpha_{i} G_{4, i}^{X}.
	\end{split}
\end{equation}

Step 9: Evaluate current $E_{X,i}$ and $E_{Z,i}$. If both values are below the set threshold, exit the feedback program; otherwise, continue with steps 2--9.

\section{DETAILED EXPERIMENTAL RESULTS}\label{appendix-DECOY}

Table~\ref{parameter} shows the results of QKD experiments.

\begin{table}[ht!]
	\centering
	\caption{Experimental results. $\mu~(\nu)$ is the intensity of the signal (decoy) state, $p_{\mu}~(p_{\nu})$ is the transmission probability of the signal (decoy) state, and $p_{z}~(p_{x})$ is the probability of the Z~(X) basis emission. The $n_{z,\mu}~(n_{z,\nu})$ is the raw count of the signal state (decoy state) measured under the Z-basis, $n_{x,\mu}~(n_{x,\nu}) $ is the raw count measured under the X-basis for the signal state (decoy state), $m_{z,\mu}~(m_{z,\nu})$ is the error count measured under the Z-basis for the signal state (decoy state), $ m_{x,\mu}~(m_{x,\nu})$ is the error count of the signal state (decoy state) measured under the X-basis, $\phi _{z}^{u} $ is the upper bound of the phase error rate, QBER is the average error rate, $s_{z,1}^{l}$ is the lower bound for measuring single-photon events under the Z-basis, and SKR is the final secure key rate.}
	\renewcommand\arraystretch{1.3}
	\tabcolsep=0.003cm
	\scalebox{1}
	{\begin{tabular}{cccccc}
			\hline \hline
			\multicolumn{1}{c}{~}   &~~~~~&       	\multicolumn{4}{c}{Channel}     
			\\  \cline{3-6} 
			Parameters	~&&      $25$ km&$50$ km&$75$ km&$100$ km\\
			\cline{3-6} 
			~&&      $4.97$ dB&$9.73$ dB&$14.22$ dB&$18.68$ dB\\
			\hline 
			$\mu$&&$0.679$&$0.654$&$0.626$&$0.569$\\
			$\nu$&&$0.127$&$0.151$ &$0.176$&$0.185$\\
			$P_{\mu}$&&$0.859$&$0.809$&$0.726$&$0.598$\\
			$P_{\nu}$&&$0.141$&$0.191$&$0.274$&$0.402$\\
			$P_{z}$&&$0.960$&$0.943$&$0.907$&$0.761$\\
			$P_{x}$&&$0.040$&$0.057$&$0.093$&$0.239$\\
			$n_{z,\mu}$&&$3.08\times{10^7}$&$8.56\times{10^6}$&$2.23\times{10^6}$&$4.53\times{10^5}$\\
			$n_{z,\nu}$&&$9.57\times{10^5}$&$4.75\times{10^5}$&$2.31\times{10^5}$&$1.05\times{10^5}$\\
			$n_{x,\mu}$&&$1.09\times{10^6}$&$5.03\times{10^5}$&$2.22\times{10^5}$&$1.43\times{10^5}$\\
			$n_{x,\nu}$&&$2.91\times{10^4}$&$2.69\times{10^4}$&$2.61\times{10^4}$&$3.53\times{10^4}$\\
			$m_{z,\mu}$&&$1.59\times{10^5}$&$4.70\times{10^4}$&$1.90\times{10^4}$&$9.90\times{10^3}$\\
			$m_{z,\nu}$&&$8.04\times{10^3}$&$4.48\times{10^3}$&$4.73\times{10^3}$&$4.68\times{10^3}$\\
			$m_{x,\mu}$&&$8.14\times{10^3}$&$4.91\times{10^3}$&$3.00\times{10^3}$&$4.84\times{10^3}$\\
			$m_{x,\nu}$&&$8.84\times{10^2}$&$8.95\times{10^2}$&$1.03\times{10^3}$&$2.92\times{10^3}$\\
			$\phi _{z}^{u} $&&$3.37\times{10^{-2}}$&$3.36\times{10^{-2}}$&$3.25\times{10^{-2}}$&$7.62\times{10^{-2}}$\\
			$QBER$&&$5.27\times{10^{-3}}$&$5.70\times{10^{-3}}$&$9.66\times{10^{-3}}$&$2.61\times{10^{-2}}$\\
			$s_{z,1}^{l} $&&$1.48\times{10^7}$&$4.26\times{10^6}$&$1.08\times{10^6}$&$2.64\times{10^5}$\\
			$SKR$(bps) &&$4.94\times{10^4}$&$1.41\times{10^4}$&$3.15\times{10^3}$&$2.40\times{10^2}$\\
			\hline \hline 
	\end{tabular}}
	\label{parameter}
\end{table}

\clearpage

\bibliography{Ref}

\end{document}